\documentclass[conference]{IEEEtran}
\IEEEoverridecommandlockouts

\usepackage[noadjust]{cite}

\usepackage{amsmath,amssymb,amsfonts}
\usepackage{algorithmic}
\usepackage{graphicx}
\usepackage{subfigure} 
\usepackage{textcomp}
\usepackage{xcolor}
\usepackage{soul}
\usepackage[capitalize,nameinlink]{cleveref}
\crefname{figure}{Fig.}{Figs.}
\crefname{equation}{}{}

\usepackage{titlesec}
\titlespacing{\section}{0pt}{0.5\baselineskip}{0.3\baselineskip}
\titlespacing{\subsection}{0pt}{0.4\baselineskip}{0.2\baselineskip}
\titlespacing{\subsubsection}{0pt}{0.3\baselineskip}{0.1\baselineskip}

\usepackage{etoolbox}
\AtBeginEnvironment{align}{\vspace{-0.2\baselineskip}} 
\AfterEndEnvironment{align}{\vspace{-0.2\baselineskip}}

\def\BibTeX{{\rm B\kern-.05em{\sc i\kern-.025em b}\kern-.08em
    T\kern-.1667em\lower.7ex\hbox{E}\kern-.125emX}}
\begin{document}

\title{DoDTrack: Indoor Mobile Devices Tracking via Difference-of-Doppler \vspace{-8pt}}

\author{
    \IEEEauthorblockN{Jingwen Zhang\IEEEauthorrefmark{1}, Chunxi Chen\IEEEauthorrefmark{1}, Chao Yu\IEEEauthorrefmark{2}, Zezhong Zhang\IEEEauthorrefmark{3}, Rui Wang\IEEEauthorrefmark{2}}
    \IEEEauthorblockA{\IEEEauthorrefmark{1} College of Semiconductors, Southern University of Science and Technology, Shenzhen, China}
    \IEEEauthorblockA{\IEEEauthorrefmark{2} Department of Electronic and Electrical Engineering, Southern University of Science and Technology, Shenzhen, China}
    \IEEEauthorblockA{\IEEEauthorrefmark{3} The Future Network of Intelligence Institute (FNii), The Chinese University of Hong Kong, Shenzhen, China}
    \IEEEauthorblockN{Email: \{zhangjingwen2024, chencx2024, 12431241\}@mail.sustech.edu.cn,
    \\
    zhangzezhong@cuhk.edu.cn, wangr@sustech.edu.cn}
    \vspace{-30pt}
}

\maketitle

\begin{abstract}
In this paper, the Doppler frequency shift (DFS) is exploited as the only sensing parameter for low-cost indoor mobile device tracking. The existing trajectory tracking methods via DFS of Wi-Fi systems often require the knowledge of the starting position or additional information, like angle-of-arrival (AoA) and time-of-flight (ToF), to recover the trajectory of a moving target. This paper proposes the DoDTrack, a novel Difference-of-Doppler (DoD)-based tracking system, to track an active mobile device using a single receiver with distributed antennas. By comparing the signals received at the distributed receive antennas, which share the oscillator, the DoDs among the antennas can be detected robustly. Then, the reconstruction of the trajectory without prior knowledge of the trajectory starting position can be formulated as a minimum mean square error (MMSE) problem, which can be solved via alternating optimization. Particularly, the starting position and the trajectory shape are updated alternately in the proposed algorithm. In performance validation, we implemented the proposed DoDTrack design on a USRP-X310 platform, and assessed its estimation accuracy with various trajectory shapes in an indoor environment. Experimental results demonstrate that DoDTrack achieves a median tracking error of 0.34 m within a 6 m $\times$ 6 m sensing area, offering a high-precision and low-cost solution for active device tracking.
\end{abstract}


\section{Introduction}
With the rapid advancement of ubiquitous computing and the Internet of Things (IoT) technologies, precise trajectory tracking of indoor mobile devices has become a core requirement of numerous location-based services, including smart buildings, industrial automation, etc. Compared to the mature global navigation satellite systems (GNSS) used in outdoor environments, indoor navigation presents long-standing challenges in achieving high-precision, low-cost trajectory tracking.

Researchers from academia and industry have proposed a number of indoor positioning and tracking solutions for active mobile devices via communication signals. Existing approaches can be roughly categorized into two main types: 1) data-driven methods based on fingerprints, such as received signal strength indicators \cite{laafu,wideep} or channel state information \cite{deepfi,cifi}; and 2) model-driven approaches based on geometric relationships, such as time-of-flight (ToF) \cite{tonetrack,sifi,rtt} or angle-of-arrival (AoA) \cite{monoloco,aoa}. On the one hand, the methods of the first type typically require substantial offline training data, which are thereby limited in some practical application scenarios. On the other hand, the methods of the second type depend on strict time synchronization between transceivers or a sufficiently large signal bandwidth, leading to substantial hardware costs and system complexity.

In recent years, the DFS has gained increasing attention as a physical-layer feature rich in dynamic environments. Its fundamental limitation on tracking, however, is that the DFS measurements alone can only reconstruct relative displacement of the moving device. To recover an accurate trajectory, an estimate of the starting position is indispensable. Consequently, existing approaches on mobile device (or passive target) tracking via DFS often require either a known starting position \cite{witraj} or complementary information like AoA \cite{indotrack,widfs} or ToF \cite{widar2}. Moreover, prior works leveraging DFS have predominantly focused on tracking passive targets (e.g., humans). There is still no work on tracking active wireless devices via DFS only.

To address this issue, this paper proposes a Difference-of-Doppler-Based Tracking system, namely DoDTrack, which can be implemented in a cellular base station (BS) or a Wi-Fi access point (AP) with distributed antennas in uplink transmission. With the knowledge of the
position of the receive antennas, the DoDTrack can reconstruct the trajectory of a moving active device via the DoDs of its signals among the receive antennas. The main contributions of this work are summarized as follows:
\begin{itemize}
\item To the best of our knowledge, this is the first effort on the trajectory reconstruction of active mobile device with DoD only.

\item The trajectory reconstruction is formulated as an MMSE problem, minimizing the gap between the observed DoDs and the analytical ones calculated from the trajectory. Moreover, a low-complexity solution algorithm based on alternating optimization is proposed.

\item To demonstrate the effectiveness of the proposed DoDTrack method, an experimental platform is developed for practical performance evaluation. Experimental results show that the DoDTrack achieves a median tracking error of 0.34 m in a 6 m $\times$ 6 m sensing area for an active Wi-Fi device.
\end{itemize}

The remainder of this paper is structured as follows. Section \ref{system-model} establishes the motion and signal models. Section \ref{DoDTrack} presents the DoDTrack design, including the DoD estimation procedures and the problem formulation. Section \ref{section: solution} proposes the trajectory reconstruction solution and starting position optimization method. Experimental results are provided in Section \ref{experiments}, followed by conclusions in Section \ref{conclusion}.

\section{System Model}
\label{system-model}
\begin{figure}[htbp]
\vspace{-10pt}
    \centering
    \includegraphics[width=0.85\columnwidth]{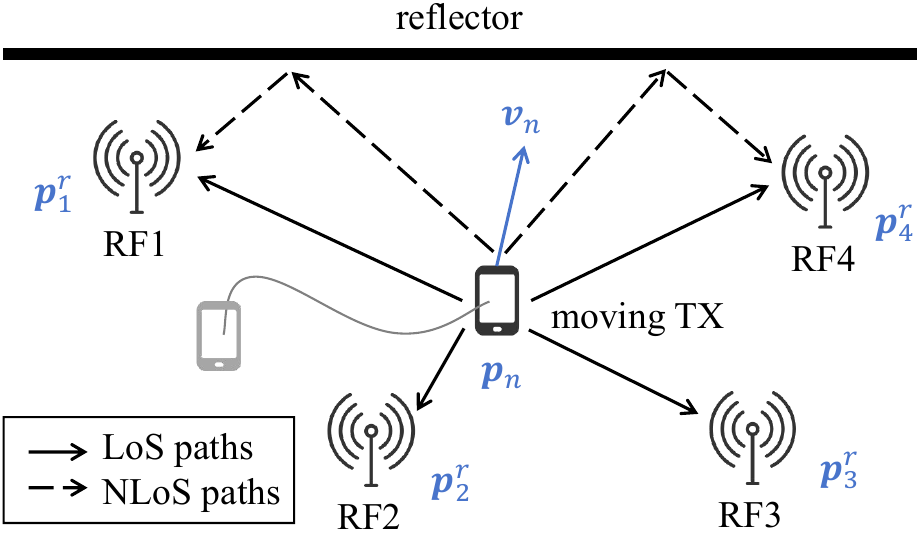}
    \caption{Illustration of mobile device tracking using distributed receive antennas.}
    \label{fig: scenario}
    \vspace{-5pt}
\end{figure}

In this work, we propose an integrated sensing and communication (ISAC) system with distributed antennas for the tracking of active mobile devices. As illustrated in \cref{fig: scenario}, the tracking system consists of one receiver with $M$ spatially separated radio frequency (RF) chains, each connected to a receive antenna, denoted as antenna $1, \ldots, M$, respectively. The receiver can be a dedicated sensing receiver, or a BS / AP in its uplink transmission mode. The position of each receive antenna is fixed and known. When the transmitter (it can be a Wi-Fi, cellular, or Bluetooth user) moves, different Doppler frequencies are induced on each RF chain. Due to hardware imperfections, there exist unavoidable sampling frequency offset (SFO) and carrier frequency offset (CFO) between the transmitter and the receiver, making it extremely challenging to estimate the Doppler frequency on each individual RF chain. Fortunately, since all RF chains share the same oscillator at the receiver, the CFOs and SFOs in their receive signals are the same. This implies that the DoDs across multiple receive antennas are detectable without the cooperation of the transmitter. Hence, in this paper, the DoDs among at least 4 receive antennas are exploited to reconstruct the trajectory of a moving target, after observing the motion for a short period. In the reconstruction, the receiver has no prior knowledge of the starting position of the moving target.

\subsection{Motion Model}
The position of the $m$-th receive antenna ($m=1,...,M$) is denoted as $\boldsymbol{p}_m^r=[x_m^r, y_m^r]^\mathsf{T}$. Given a total of $K$ observation intervals, each of duration $\Delta t$, the position and velocity of the transmitter at the beginning of the $k$-th observation interval are denoted as $\boldsymbol{p}_k = [x_k, y_k]^{\mathsf{T}}$ and $\boldsymbol{v}_k = [v^x_k, v^y_k]^{\mathsf{T}}$, respectively. When the observation interval is sufficiently small, i.e., $\Delta t \to 0$, the motion of the transmitter can be approximated by a linear model, given as\vspace{-5pt}
\begin{equation}
    \boldsymbol{p}_{k+1} = \boldsymbol{p}_k + \boldsymbol{v}_k \Delta t.
    \label{eq: motion model}  
\end{equation}

The Doppler frequency is determined by the relative motion and positions between the transmitter and the receive antenna. For the $m$-th antenna, the Doppler frequency at the $k$-th observation interval is\vspace{-5pt}
\begin{equation}
    f_m^d(k)=\dfrac{f_c}{c}
    \left (\frac{\boldsymbol{p}_m^r-\boldsymbol{p}_k}{\|\boldsymbol{p}_m^r-\boldsymbol{p}_k\|}\right )^\mathsf{T}\boldsymbol{v}_k,
    \label{eq: Doppler}    
\end{equation}
where $f_c$ is the carrier frequency, $c$ is the speed of light and $\|\cdot\|$ denotes the norm of the vector. Accordingly, the Difference-of-Doppler frequency (DoD) between the $m$-th and $n$-th antennas ($m \neq n$) is given by
\begin{align}
    \Delta f_{m,n}^d(k) &= f_m^d(k)- f_n^d(k) \nonumber \\
    &= \frac{f_c}{c}\left(\frac{\boldsymbol{p}_m^r-\boldsymbol p_k}{\|\boldsymbol{p}_m^r-\boldsymbol p_k\|}-\frac{\boldsymbol{p}_n^r-\boldsymbol p_k}{\|\boldsymbol{p}_n^r-\boldsymbol p_k\|}\right)^\mathsf{T}
    \boldsymbol v_k.
    \label{eq: Doppler_Difference}
\end{align}

\vspace{-5pt}
\subsection{Signal Model}
Let $x(t)$ be the transmitted signal. The signal received by the $m$-th antenna can be expressed as the combination of Line-of-Sight (LoS) component $y_m^{\mathrm{LoS}}(t)$, Non-Line-of-Sight (NLoS) components $y_m^{\mathrm{NLoS}}(t)$ and noise $n_m(t)$, given by
\begin{equation}
    y_m(t)=e^{-j2\pi(\int_0^t f_{\text{off}}(t)dt) }[y_m^{\mathrm{LoS}}(t)+y_m^{\mathrm{NLoS}}(t)]+n_m(t) , 
    \label{eq: signal_model}
\end{equation}
with 
\begin{align*}
    &y_m^{\mathrm{LoS}}(t)=\alpha_{m,1}(t)x(t-\tau _{m,1}(t))e^{-j2\pi (\int_0^t f_{m,1}(\tau)d\tau)+\varphi_{m,1}},\\
    &y_m^{\mathrm{NLoS}}(t)=\sum_{l=2}^{L_m}\alpha_{m,l}(t)x(t-\tau _{m,l}(t))e^{-j2\pi (\int_0^t f_{m,l}(\tau)d\tau)+\varphi_{m,l}},
\end{align*}
where $f_{\text{off}}(t)$ denotes the frequency offset between the transmitter and receiver, including both CFO and SFO. $L_m$ represents the number of propagation paths. The set $\{\alpha_{m,l}(t), \tau_{m,l}(t), f_{m,l}(t), \varphi_{m,l}\}$ represents the complex gain, propagation delay, Doppler shift, and initial phase of the $l$-th path, respectively, with $l = 1$ corresponding to the LoS path. Note that $f_m^d(k)$ defined in \cref{eq: Doppler} is the Doppler frequency of the LoS path from the transmitter to the $m$-th antenna at the time $k\Delta t$, we have $f_{m,1}(k\Delta t)=f_m^d(k)$.

\section{Difference-of-Doppler-Based Tracking}\label{DoDTrack}
In this section, we present the framework of the proposed DoDTrack approach. The DoD detection process is first described. Then the tracking problem is formulated as an optimization problem to minimize the gap between the analytical and measured DoDs.

\subsection{DoD Detection}
\begin{figure*}[htbp] 
\vspace{-5pt}
 	\centering
 	\begin{align}
 	    y_{m,n}(t)&=y_m(t)y_n^{*}(t)=y_m^{\mathrm{LoS}}(t)[y_n^{\mathrm{LoS}}(t)]^*+y_m^{\mathrm{LoS}}(t)[y_n^{\mathrm{NLoS}}(t)]^*+y_m^{\mathrm{NLoS}}(t)[y_n^{\mathrm{LoS}}(t)]^*+y_m^{\mathrm{NLoS}}(t)[y_n^{\mathrm{NLoS}}(t)]^*+n_{m,n}(t) \nonumber \\
 	    &=\alpha_{m,n}^{1,1}(t)|x(t)|^2e^{-j2\pi (\int_0^t \Delta f_{m,n}^{1,1}(\tau)d\tau)+\Delta\varphi_{m,n}^{1,1}}+
 	     \sum_{k=2}^{L_n}\alpha_{m,n}^{1,k}(t)|x(t)|^2e^{-j2\pi (\int_0^t \Delta f_{m,n}^{1,k}(\tau)d\tau)+\Delta\varphi_{m,n}^{1,k}}+ \nonumber \\
 	    &\sum_{l=2}^{L_m}\alpha_{m,n}^{l,1}(t)|x(t)|^2e^{-j2\pi (\int_0^t \Delta f_{m,n}^{l,1}(\tau)d\tau)+\Delta\varphi_{m,n}^{l,1}}+
 	    \sum_{l=2}^{L_m}\sum_{k=2}^{L_n}\alpha_{m,n}^{l,k}(t)|x(t)|^2e^{-j2\pi (\int_0^t \Delta f_{m,n}^{l,k}(\tau)d\tau)+\Delta\varphi_{m,n}^{l,k}}+n_{m,n}(t)
 	    \label{eq: conj_multi}   
 	\end{align}
 	\vspace{-15pt}
\end{figure*}

As aforementioned, due to the presence of frequency offset and multi-path effects, the target Doppler frequency cannot be directly obtained based on the raw received signals. To address this issue, we first perform conjugate multiplication on raw received signals from different antennas to eliminate the effect of the frequency offset. Subsequently, the Short-Time Fourier Transform (STFT) is applied to extract the target DoD information.

In order to maintain a low cost, the proposed DoDTrack relies only on the DoD measurements, instead of distance measurements. Specifically, a narrowband portion of the received signal is sufficient for the trajectory reconstruction, reducing the sampling rate and computation complexity for sensing. The misalignment of $\tau_{m,l}(t)$ from different paths is negligible. Therefore, we neglect the effect of $\tau_{m,l}(t)$ in the signal model. As a result, given the received signals at the $m$-th and $n$-th antennas, the frequency offset can be removed by conjugate multiplication as presented in \cref{eq: conj_multi}, where\vspace{-3pt}
\begin{align*}
\vspace{-3pt}
     \alpha_{m,n}^{l,k}(t)=&\alpha_{m,l}(t)\alpha_{n,k}^*(t),\quad \Delta\varphi_{m,n}^{l,k}=\varphi_{m,l}-\varphi_{n,k},\\
    &\Delta f_{m,n}^{l,k}(\tau)=f_{m,l}(\tau)-f_{n,k}(\tau),
\end{align*}
$\{\cdot\}^*$ denotes the conjugate operation, and $n_{m,n}(t)$ denotes the merged noise. Note that the noise can be neglected since its power is significantly lower than that of the multipath signals.

Applying STFT to the signal yields the time-frequency spectrum as
\vspace{-5pt}
\begin{equation}
\vspace{-3pt}
    Y_{m,n}(t,\Delta f)=\int_{T_w}y_{m,n}(t)w(\tau-t)e^{j2\pi\Delta f\tau}d\tau,
    \label{eq: STFT}
\end{equation}
where $w(\cdot)$ denotes a window function and $T_w$ is the window length. The time-frequency spectrum represents the strength of each frequency component. The signal $y_{m,n}(t)$ is the superposition of all multipath Doppler difference components. Fortunately, their amplitudes differ significantly because LoS signals are usually much stronger than NLoS signals (i.e., $\alpha_{m,1}>\alpha_{m,l}$). Consequently, the peak in the time-frequency spectrum coincides with the LoS component with amplitude $\alpha_{1,1}$ and frequency $\Delta f_{m,n}^{1,1}(t)$. Therefore, the DoD between received signals at antennas $m$ and $n$ can be estimated by detecting the frequency of the spectral peak, given as\vspace{-3pt}
\begin{equation}
    \Delta\tilde{f}^d_{m,n}(t)=\mathop{\arg\max}\limits_{\Delta f}\left|Y_{m,n}(t,\Delta f)\right|.
    \label{eq: DoD_extract}
    \vspace{-3pt}
\end{equation}
Note that the analytical DoD given the trajectory can be calculated according to \cref{eq: Doppler_Difference}, the difference between them is due to the estimation error and the trajectory reconstruction error.

\subsection{Problem Formulation}
Given the analytical DoD model in \cref{eq: Doppler_Difference} and the detected DoD in \cref{eq: DoD_extract}, we formulate the device tracking task as an MMSE problem, which is to find the optimal trajectory of the transmitter that minimizes the difference between the analytical and detected DoD.

With $M$ antennas,  $\mathit{C}_M^2=M!/[(M-2)!2!]$ distinct DoDs can be obtained at each observation interval. The DoD observation vector at the $k$-th observation interval can be represented as
\begin{equation}
\vspace{-3pt}
    \!\!\tilde{\boldsymbol{z}}_k\!=\![\Delta\! \tilde{f}^d_{1,2}\!(k),...,\Delta\! \tilde{f}^d_{1,M}\!(k),\Delta\! \tilde{f}^d_{2,3}\!(k),...,\Delta\! \tilde{f}^d_{M\!-\!1,M}\!(k)]^\mathsf{T}.
    \label{eq: dod_vector}
\end{equation}
The DoD observations across $K$ observation intervals can be expressed as the following observation matrix
\begin{equation}
\vspace{-3pt}
    \widetilde{\mathbf{Z}}=[\tilde{\boldsymbol{z}}_1,\tilde{\boldsymbol{z}}_2,...,\tilde{\boldsymbol{z}}_K].
\end{equation}
Moreover, according to \cref{eq: motion model}, the trajectory of TX can be expressed equivalently as the combination of the starting position $\boldsymbol{p}_1$ and the subsequent velocities, denoted as\vspace{-3pt}
\begin{equation}
\vspace{-3pt}
    \mathbf{P}=[\boldsymbol{p}_1,\boldsymbol{v}_1,...,\boldsymbol{v}_{K-1}],
\end{equation}
the corresponding analytical DoD matrix can be expressed as \vspace{-8pt}
\begin{equation}
\vspace{-3pt}
    \mathbf{Z}(\mathbf{P})=[\boldsymbol{z}_1,\boldsymbol{z}_2,...,\boldsymbol{z}_K],
\end{equation}
where $\boldsymbol{z}_k$ is the analytical DoD vector\vspace{-2pt}
\begin{equation*}
\vspace{-2pt}
    \boldsymbol{z}_k=\![\Delta\! f^d_{1,2}\!(k),..., \Delta\!f^d_{m,n}\!(k),...,\Delta\! f^d_{M\!-\!1,M}\!(k)]^\mathsf{T},
\end{equation*}
and $\Delta\!f^d_{m,n}\!(k)$ can be calculated according to \cref{eq: Doppler_Difference}.
Therefore, the normalized MSE between the above two DoD matrix is defined as
\begin{align}
    &g(\mathbf{P},\widetilde{\mathbf{Z}})=\frac{1}{\mathit{C}_M^2K}\|\widetilde{\mathbf{Z}}-\mathbf{Z}(\mathbf{P})\|_\mathrm{F}^2 \nonumber \\
    &=\frac{1}{\mathit{C}_M^2K}\sum_{k=1}^K\sum_{i=1}^{M-1}\sum_{j=i+1}^M[\Delta\tilde{f}^d_{i,j}(k)-\Delta f^d_{i,j}(k)]^2,
\end{align}
where $\|\cdot\|_\mathrm{F}^2$ denotes the Frobenius norm. As a result, the trajectory reconstruction problem can be expressed as
\begin{align}
    \mathcal{P}_1 : \quad \mathbf{P}^*&=\mathop{\arg\min}_{\mathbf{P}}g(\mathbf{P},\widetilde{\mathbf{Z}}).
    \label{eq: problem1}
\end{align}

\vspace{-5pt}
\section{Low-Complexity Solution}
\label{section: solution}
Since the denominator in \cref{eq: Doppler_Difference} involves L2-Norm, the optimization in \cref{eq: problem1} is a non-convex optimization problem. Directly solving it usually involves substantial computational effort due to the large number of optimization variables. In this section, a low-complexity solution is proposed, where the optimization of trajectory is divided into two iterative steps: optimize the shape of the trajectory $[\boldsymbol{v}_1,...,\boldsymbol{v}_{K-1}]$ given the starting position $\boldsymbol{p}_1$, and optimize the starting position given the shape of the trajectory. We can use the Extended Kalman Filter (EKF) in the shaping of the trajectory in the first step. Moreover, to avoid the trapping at local optimal, the above iteration can start from multiple initial solutions of the starting position.

Particularly, let $    \mathcal{J}^0=\{\boldsymbol{p}_{1,j}^0|j=1,2,...,J\}$ be the set of initial solutions of the starting position. The two sub-problems in the $i$-th iteration originated from the $j$-th initial solution in the set $\mathcal{J}$ can be expressed as
\begin{align}
    &\mathcal{P}_{1-1}^{(i,j)} : \quad [\boldsymbol{p}_{1,j}^{i-1},\boldsymbol{v}_{1,j}^i,...,\boldsymbol{v}_{K-1,j}^i]=f_{\mathrm{EKF}}(\boldsymbol{p}_{1,j}^{i-1},\widetilde{\mathbf{Z}}),\nonumber \\ 
    &\mathcal{P}_{1-2}^{(i,j)} : \quad \boldsymbol{p}_{1,j}^i=\mathop{\arg\min}_{\boldsymbol{p}_1}g([\boldsymbol{p}_1,\boldsymbol{v}_{1,j}^i,...,\boldsymbol{v}_{K-1,j}^i],\widetilde{\mathbf{Z}}),
    \label{eq: sub-problems}
\end{align}
where $f_{\mathrm{EKF}}(\cdot)$ is the operator of EKF.\vspace{5pt}

In the following, the EKF-based trajectory shaping and the optimization of starting position are elaborated respectively.

\subsection{Trajectory Reconstruction using EKF}
\label{sec: solution_EKF}
In this part, the EKF-based trajectory reconstruction given the starting position $\boldsymbol{p}_1$ and the DoD observations $\widetilde{\mathbf{Z}}$ is provided.

The main steps of the EKF are state prediction and state update. Define the state vector as $\boldsymbol s_k=[x_k,y_k,v^x_k,v^y_k]^\mathsf{T}=[\boldsymbol p_k^\mathsf{T},\boldsymbol v_k^\mathsf{T}]^\mathsf{T}$, the state transition function can be expressed as \vspace{-3pt}
\begin{equation}
    \boldsymbol s_k=\mathbf{A}\boldsymbol s_{k-1}+\boldsymbol w_k, 
    \label{eq: state-transition}
    \vspace{-3pt}
\end{equation}
where
\begin{equation*}
    \mathbf{A}=
    \begin{bmatrix}
    1 & 0 & \Delta t & 0\\
    0 & 1 & 0 & \Delta t\\
    0 & 0 & 1 & 0\\
    0 & 0 & 0 & 1
    \end{bmatrix}
\end{equation*}
is the state transition matrix, $\boldsymbol{w}_k\sim \mathcal{N}(0,\mathbf{W})$ is the process noise and $\mathbf{W}$ is the covariance matrix of the process noise .

Moreover, the DoD observations $\tilde{\boldsymbol z}_k$ is incorporated as the measurement vector in the EKF process.  The measurement equation can be expressed as\vspace{-3pt}
\begin{equation}
\vspace{-5pt}
    \tilde{\boldsymbol z}_k=\boldsymbol h(\boldsymbol s_k)+\boldsymbol u_k,
    \label{eq: observation_eqation}
\end{equation}
where
\begin{equation}
    \boldsymbol{h}(\boldsymbol{s}_k)=    \begin{bmatrix}
    \frac{f_c}{c}(\frac{\boldsymbol{p}_1^r-\boldsymbol p_k}{\|\boldsymbol{p}_1^r-\boldsymbol p_k\|}-\frac{\boldsymbol{p}_2^r-\boldsymbol p_k}{\|\boldsymbol{p}_2^r-\boldsymbol p_k\|})^\mathsf{T}\boldsymbol v_k\\
    \vdots\\
    \frac{f_c}{c}(\frac{\boldsymbol{p}_{M-1}^r-\boldsymbol p_k}{\|\boldsymbol{p}_{M-1}^r-\boldsymbol p_k\|}-\frac{\boldsymbol{p}_{M}^r-\boldsymbol p_k}{\|\boldsymbol{p}_{M}^r-\boldsymbol p_k\|})^\mathsf{T}\boldsymbol v_k
    \end{bmatrix}
    \label{eq: observation_function}
\end{equation}
is the nonlinear measurement function,  $\boldsymbol{u}_k\sim \mathcal{N}(0,\mathbf{U})$ is the measurement noise and $\mathbf{U}$ is the covariance matrix.

Then a complete trajectory estimate can be obtained via iteratively predicting and updating the state vector, a complete trajectory estimate can ultimately be obtained. Let $\boldsymbol{s}_{k|k-1}$ and $\mathbf{C}_{k|k-1}$ denote prior state vector and its corresponding covariance matrix, and $\boldsymbol{s}_{k|k}$ and $\mathbf{C}_{k|k}$ denote the posterior ones based on latest observations. The formula for the prediction phase can be expressed as\vspace{-5pt}
\begin{equation}
\vspace{-4pt}
    \begin{cases}
        \boldsymbol s_{k|k-1}=\mathbf{A}\boldsymbol s_{k-1|k-1}\\
        \mathbf{C}_{k|k-1}=\mathbf{A}\mathbf{C}_{k-1|k-1}\mathbf{A}^\mathsf{T}+\mathbf{W}
    \end{cases},
    \label{eq: EKF_predict}
\end{equation}
and the update formula is\vspace{-3pt}
\begin{equation}
    \begin{cases}
        \mathbf{s}_{k|k}=\mathbf{s}_{k|k-1} + \mathbf{K}_k[\tilde{\boldsymbol z}_k - \boldsymbol{h}(\boldsymbol{s}_{k|k-1})]\\
        \mathbf{C}_{k|k}=(\mathbf{I}-\mathbf{K}_k\mathbf{H}_k)\mathbf{C}_{k|k-1}
    \end{cases} ,
\end{equation}
where $\mathbf{K}_k$ is the Kalman gain given by
\begin{equation}
    \mathbf{K}_k = \mathbf{C}_{k|k-1}\mathbf{H}_k^\mathsf{T}(\mathbf{H}_k\mathbf{C}_{k|k-1}\mathbf{H}_k^\mathsf{T} + \mathbf{U})^{-1},
\end{equation}
$\mathbf{I}$ is the identity matrix and $\mathbf{H}_k$ is the Jacobian matrix of $\boldsymbol{h}(\boldsymbol{s}_k)$ given by
\begin{equation}
    \mathbf{H}_k = 
    \begin{bmatrix} 
    \begin{smallmatrix}
    \frac{\partial \Delta f^d_{1,2}(k)}{\partial x_k}&\frac{\partial \Delta f^d_{1,2}(k)}{\partial y_k}&\frac{\partial \Delta f^d_{1,2}(k)}{\partial v^x_k}&\frac{\partial \Delta  f^d_{1,2}(k)}{\partial v^y_k}\\
    \vdots&\ddots&\ddots&\vdots\\
    \frac{\partial \Delta f^d_{M-1,M}(k)}{\partial x_k}&\cdots&\cdots&\frac{\partial \Delta f^d_{M-1,M}(k)}{\partial v^y_k}
    \end{smallmatrix} 
    \end{bmatrix}.
\end{equation}
The partial derivatives with respect to $x_k$ and $v^x_k$ are given by\vspace{-5pt}
\begin{align}
\vspace{-5pt}
    &\frac{\partial \Delta f^d_{m,n}(k)}{\partial x_k}=\frac{f_c}{c}\{v^x_k[-\frac{(y_k-y_m^r)^2}{\|\boldsymbol{p}_m^r-\boldsymbol p_k\|^3}+\frac{(y_k-y_n^r)^2}{\|\boldsymbol{p}_n^r-\boldsymbol p_k\|^3}] \nonumber \\
    &+v^y_k[\frac{(x_k-x_m^r)(y_k-y_m^r)}{\|\boldsymbol{p}_m^r-\boldsymbol p_k\|^3}-\frac{(x_k-x_n^r)(y_k-y_n^r)}{\|\boldsymbol{p}_n^r-\boldsymbol p_k\|^3}]\},
\end{align}
and
\begin{equation}
    \frac{\partial \Delta f^d_{m,n}(k)}{\partial v_k^x}=\frac{f_c}{c}[\frac{(x_m^r-x_k)}{\|\boldsymbol{p}_m^r-\boldsymbol p_k\|}-\frac{(x_n^r-x_k)}{\|\boldsymbol{p}_n^r-\boldsymbol p_k\|}],
\end{equation}
while the partial derivatives with respect to $y_k$ and $v^y_k$ follow the same form.

As elaborated above, the EKF process is capable of estimating the entire trajectory given the DoD observations and an initial state $\boldsymbol{s}_{1|1}$. Although $\boldsymbol{s}_{1|1}$ contains both starting position $\boldsymbol{p}_1$ and initial velocity $\boldsymbol{v}_1$, $\boldsymbol{v}_1$ can be solved using $\boldsymbol{p}_1$ and $\tilde{\boldsymbol{z}}_1$ as shown below.

Equation \cref{eq: observation_eqation} can be rewritten as 
\vspace{-5pt}
\begin{equation}
    \tilde{\boldsymbol{z}}_1=\mathbf{B}_1 {\boldsymbol{v}}_1 + \boldsymbol{u}_1,
\end{equation}
where
\vspace{-5pt}
\begin{equation*}
    \mathbf{B}_1=
    \begin{bmatrix}
    \frac{f_c}{c}\left(\frac{\boldsymbol{p}_1^r-\boldsymbol p_1}{\|\boldsymbol{p}_1^r-\boldsymbol p_1\|}-\frac{\boldsymbol{p}_2^r-\boldsymbol p_1}{\|\boldsymbol{p}_2^r-\boldsymbol p_1\|}\right)^\mathsf{T} \\
    \vdots \\
    \frac{f_c}{c}\left(\frac{\boldsymbol{p}_{M-1}^r-\boldsymbol p_1}{\|\boldsymbol{p}_{M-1}^r-\boldsymbol p_1\|}-\frac{\boldsymbol{p}_M^r-\boldsymbol p_1}{\|\boldsymbol{p}_M^r-\boldsymbol p_1\|}\right)^\mathsf{T}
    \end{bmatrix}.
\end{equation*}
Then the initial velocity can be estimated via a least-squares solution, given as\vspace{-5pt}
\begin{equation}
    \hat{\boldsymbol{v}}_1=(\mathbf{B}_1^\mathsf{T}\mathbf{B}_1)^{-1}\mathbf{B}_1^\mathsf{T}\tilde{\boldsymbol{z}}_1.
    \label{eq: estimate_v}
\end{equation}
The solution in \cref{eq: estimate_v} holds if $\mathbf{B}_1$ has full column rank, which requires at least three antennas are arranged in a non-collinear configuration. Finally, the process of EKF can be expressed as a function of starting position $\boldsymbol{p}_1$ and DoD observations $\widetilde{\mathbf{Z}}$, to solve the sub-problem $\mathcal{P}_{1-1}^{(i,j)}$ in \cref{eq: sub-problems}.

\subsection{Starting Position Update via Gradient Descent}
\label{sec: solution_search}
In order to simplify the notation, we ignore  some superscripts and subscripts in the problem $\mathcal{P}_{1-2}^{(i,j)}$. Thus, it can be rewritten as\vspace{-3pt}
\begin{equation}
    \boldsymbol{p}_1^*=\mathop{\arg\min}_{\boldsymbol{p}_1}g([\boldsymbol{p}_1,\boldsymbol{v}_1,...,\boldsymbol{v}_{K-1}],\widetilde{\mathbf{Z}}).
    \label{eq: problem2}\vspace{-3pt}    
\end{equation}
Note that the objective function $g([\boldsymbol{p}_1,\boldsymbol{v}_1,...,\boldsymbol{v}_{K-1}],\widetilde{\mathbf{Z}})$ is differentiable with respect to $\boldsymbol{p}_1$, which allows the use of gradient descent in the optimization.

Particularly, let $\boldsymbol{u}_0$ be the initial solution of \cref{eq: problem2}. For the sub-problem $\mathcal{P}_{1-2}^{(i,j)}$, $\boldsymbol{u}_0=\boldsymbol{p}_{1,j}^{i-1}$. Then, the gradient descent on the solution can be written as 
\begin{equation}
    \boldsymbol{u}_q=\boldsymbol{u}_{q-1}-\alpha
    \frac{\partial g([\boldsymbol{u}_{q-1},\boldsymbol{v}_1,...,\boldsymbol{v}_{K-1}],\widetilde{\mathbf{Z}})}{\partial \boldsymbol{u}_{q-1}},
\end{equation}
where $q$ is the iteration index of the gradient descent and $\alpha$ is the step size. Due to the page limitation, the expression of the above gradient is omitted here. After convergence, the solution of \cref{eq: problem2} is given by $\boldsymbol{p}_1^*=\boldsymbol{u}_Q$, where $Q$ is the step number of the gradient descent.

After the convergence of the two iterative optimizations in \cref{eq: sub-problems}, let $\boldsymbol{p}_{1,j}^*$ be the final solution of the starting position originating from the initial solution $\boldsymbol{p}_{1,j}^0$, the set of all final solutions can be denoted as
\begin{equation}
    \mathcal{J}^*=\{\boldsymbol{p}_{1,j}^*|j=1,2,...,J\}.
\end{equation}
The optimal starting position is selected as the one in $\mathcal{J}^*$ that minimizes the objective function, given as
\begin{equation}
    \boldsymbol{p}_1^*=\mathop{\arg\min}_{\boldsymbol{p}_{1,j}^*\in \mathcal{J}^*}g([\boldsymbol{p}_{1,j}^*,\boldsymbol{v}_{1,j}^*,...,\boldsymbol{v}_{K-1,j}^*],\widetilde{\mathbf{Z}}),
\end{equation}
where the trajectory can be obtained from EKF as 
\begin{equation}
    [\boldsymbol{p}_{1,j}^*,\boldsymbol{v}_{1,j}^*,...,\boldsymbol{v}_{K-1,j}^*]=f_{\mathrm{EKF}}(\boldsymbol{p}_{1,j}^*,\widetilde{\mathbf{Z}}).
\end{equation}

\vspace{5pt}
\section{Experiments and Results Discussion}\label{experiments}
\subsection{Experimental Setup}\vspace{-3pt}
\begin{figure}[htbp]
\vspace{-5pt}
    \centering
    \includegraphics[width=1\columnwidth]{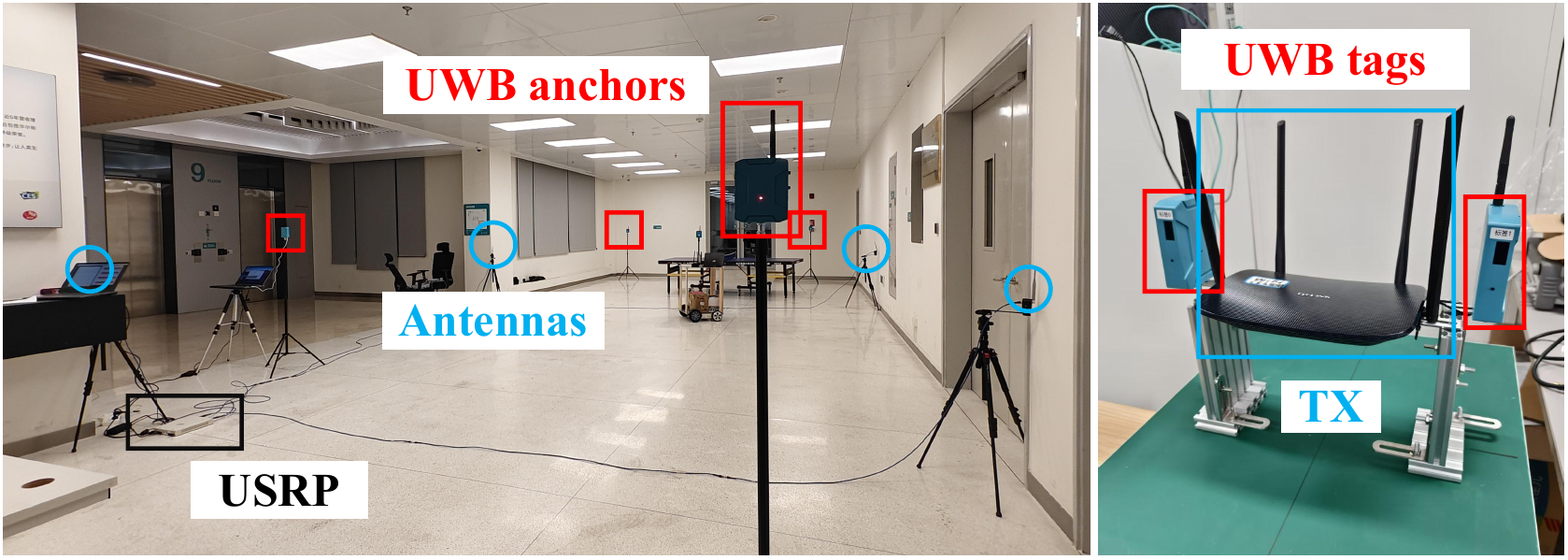}\vspace{-5pt}
    \caption{Experimental layout of DoDTrack.}
    \label{fig: layout}
    \vspace{-5pt}
\end{figure}
The experiments were conducted in an indoor environment to validate the effectiveness of DoDTrack, as shown in \cref{fig: layout}. A commercial router TP-LINK TL-XDR1520 was used as the transmitter and a four-channel USRP-X310 connected to four distributed omnidirectional antennas was deployed as the receiver. The transmitter is working on a 5.32 GHz carrier frequency with 20 MHz bandwidth, and the sampling rate of USRP receiver is 2 MHz. Note that the DoDTrack only measures the Doppler frequencies, a sampling rate of 2 MHz is sufficient. 

\begin{figure}[htbp]
    \centering
    \includegraphics[width=0.8\columnwidth]{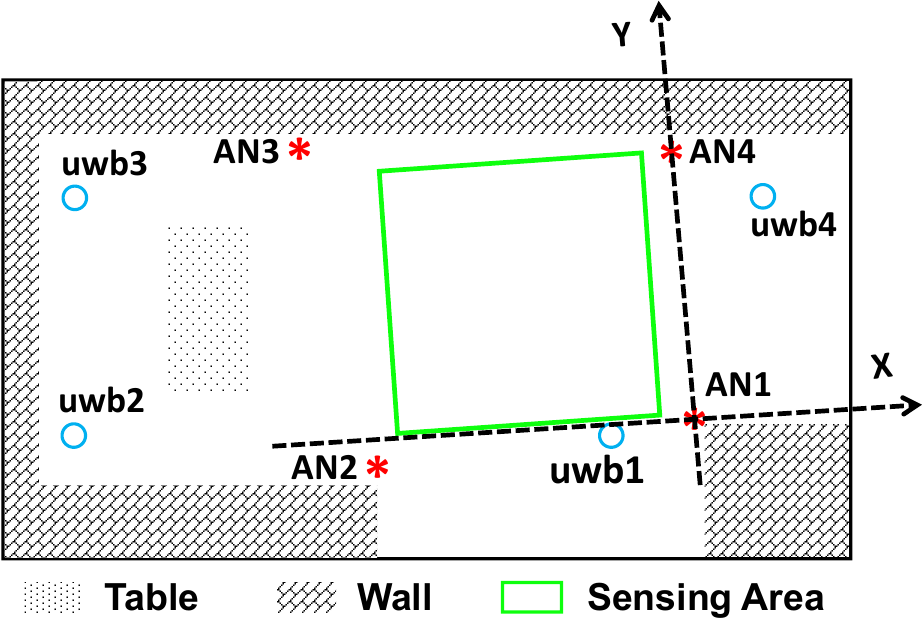}
    \caption{Top-view of the experimental sensing scenario.}
    \label{fig: floor_plan}
    \vspace{-10pt}
\end{figure}

The router was mounted on a remote-controlled mobile cart to simulate the transmitter's movement. The sensing area is approximately 6 m $\times$ 6 m, and the antenna positions are measured in advance. To record the ground truth trajectories, four UWB anchors were deployed in the room and two UWB tags were installed on the cart, which could achieve a location accuracy of approximately 10 cm. We collected 80 trajectories of different shapes, including circle, rectangle, and other random shapes. For presentation convenience, a Cartesian coordinate system is established with the antenna 1 as the origin and the antenna 4 located on the positive half-axis of the y-axis, as shown in \cref{fig: floor_plan}.

\subsection{DoD Detection}
The choice of window length in STFT is a trade-off between time resolution and frequency resolution. Specifically, a longer window provides better frequency resolution (important for accurate Doppler estimation) but worse time resolution (as it may smoothen the rapid motion changes). Given the facts that indoor motion trajectories are typically smooth and precise DoD observations are necessary in this work, we set a window length of 0.5 s, corresponding to a frequency resolution of 2 Hz.\vspace{-5pt}
\begin{figure}[htbp]
\vspace{-5pt}
    \centering
    \includegraphics[width=0.8\columnwidth]{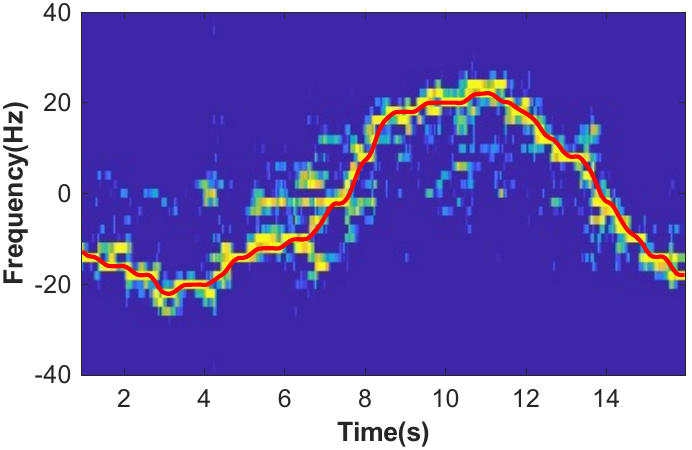}
    \vspace{-5pt}
    \caption{DoD detection based on the time-frequency spectrum.}
    \label{fig: spectrum}
\end{figure}

\cref{fig: spectrum} shows the time-frequency spectrum of DoDs between the antenna 2 and antenna 4, corresponding to the trajectory shown in \cref{fig: a}. The color ranges from blue to yellow, indicating signal strength from weak to strong, and the solid red line represents the detected DoDs. The yellow areas not covered by the red line represent noise or other strong multi-path components. It can be seen that the DoD is initially negative and then positive, resembling a sine wave, which matches the analytical DoDs of this trajectory.

\subsection{Analysis of Tracking Results}

\begin{figure}[htbp]
\centering
\subfigure[Results for a circle trajectory.]{
\includegraphics[width=0.45\linewidth]{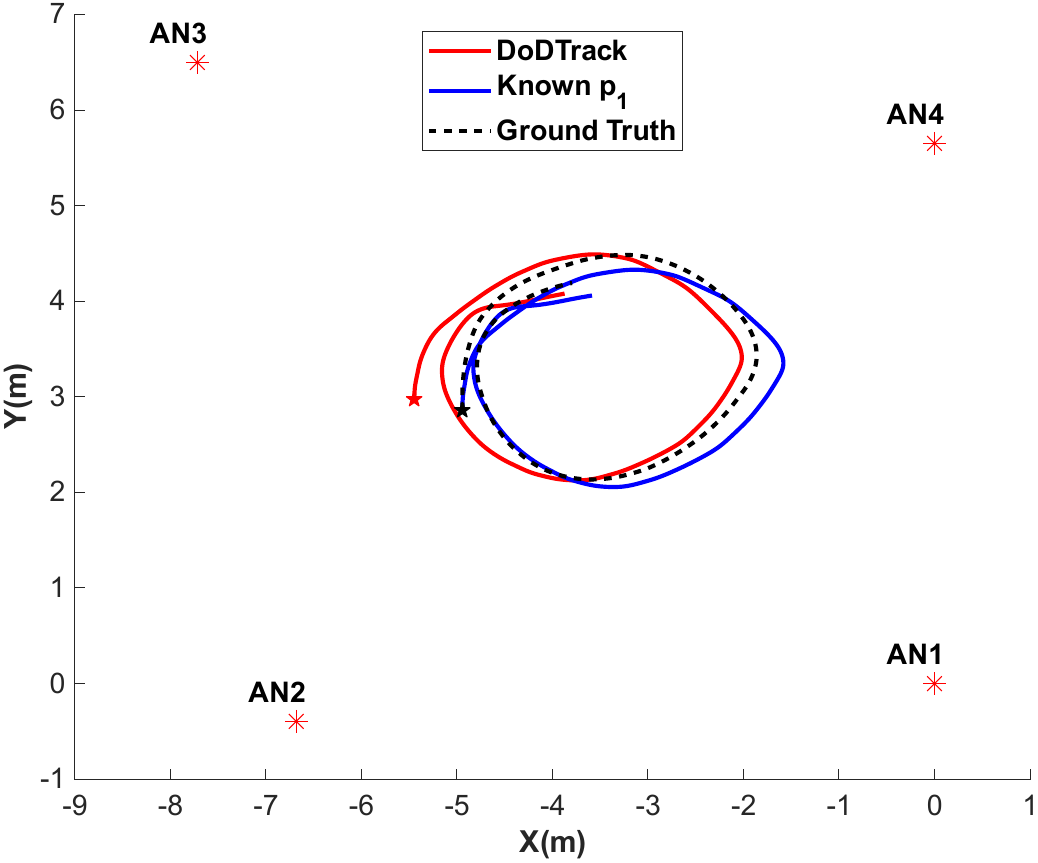}
\label{fig: a}
}
\subfigure[Results for a rectangle trajectory.]{
\includegraphics[width=0.45\linewidth]{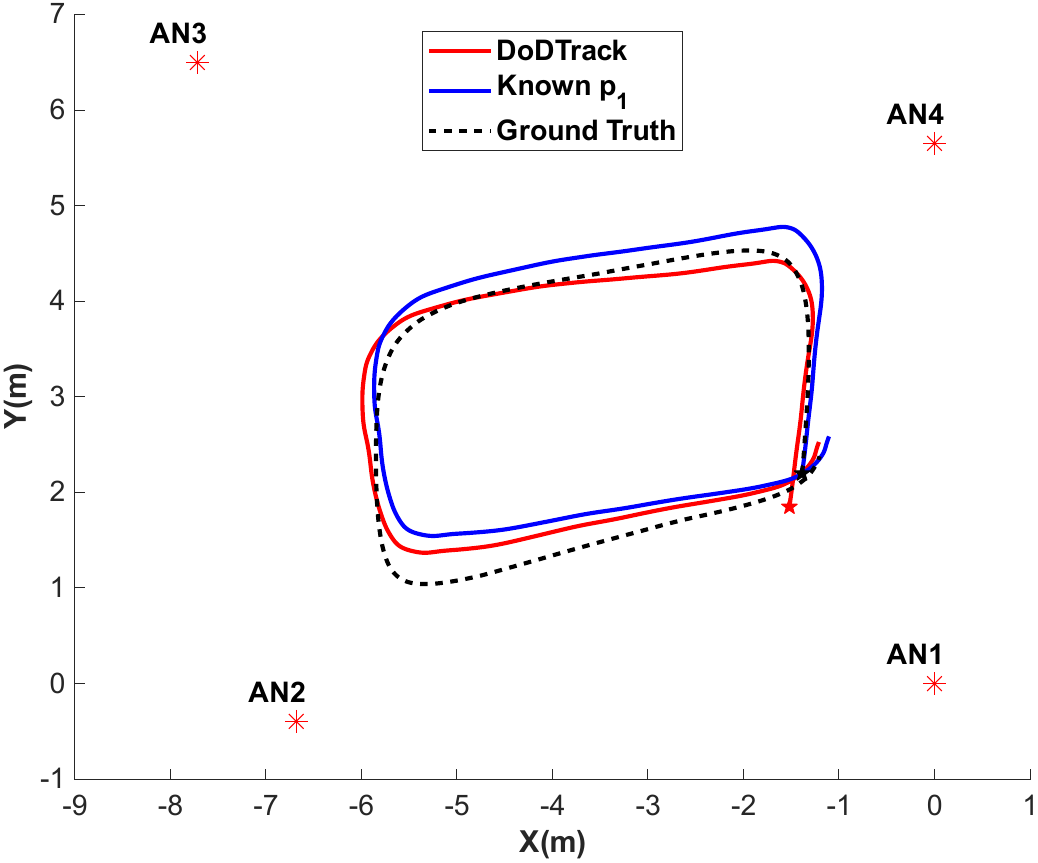}
\label{fig: b}
}

\subfigure[Results for a random trajectory.]{
\includegraphics[width=0.45\linewidth]{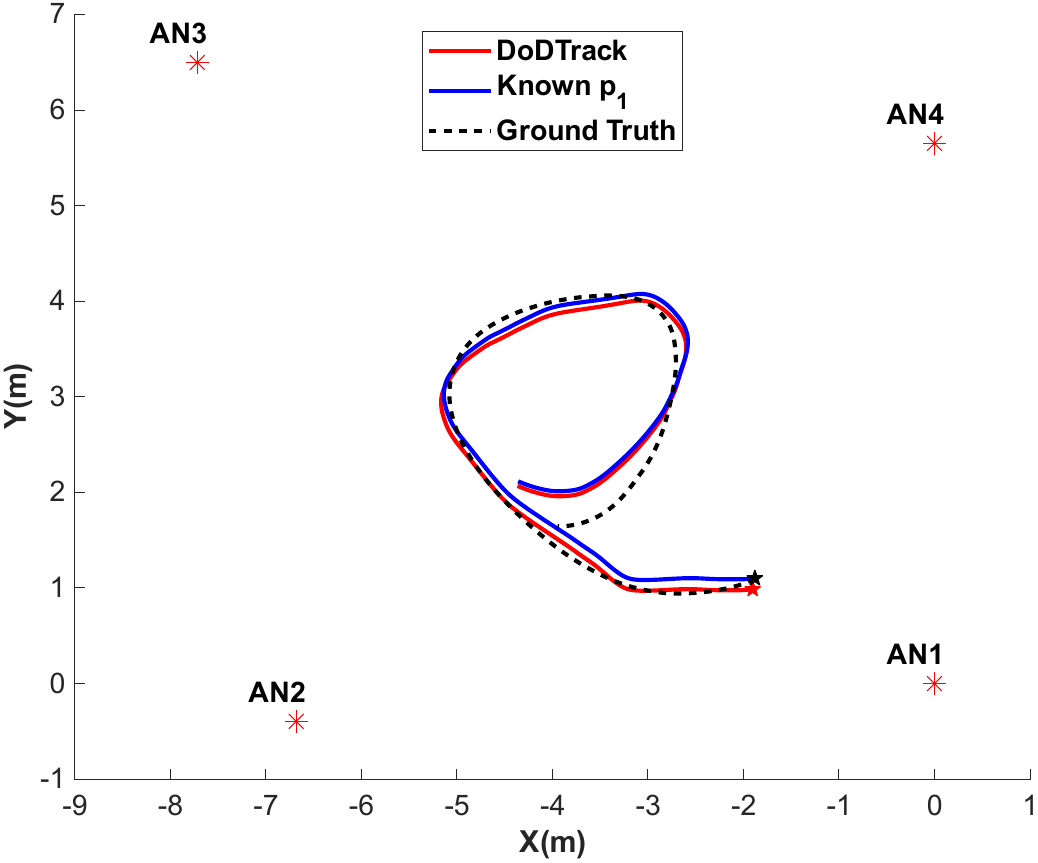}
\label{fig: c}
}
\caption{Tracking results of DoDTrack for trajectories of different shapes.}
\label{fig: results}
\end{figure}

Without loss of generality, the initial solutions of the starting position are uniformly distributed within the sensing area (hereinafter referred to as grid points). \cref{fig: results} shows the reconstruction results of DoDTrack for trajectories of different shapes when the number of grid points is $J=100$. The black dashed line represents the ground truth, and the red solid line indicates the reconstruction by DoDTrack. For comparison, we also provide the tracking results by applying EKF on the ground truth of starting position $\boldsymbol{p}_1$, which is marked in blue. The red asterisk represents the receive antennas and the pentagram denotes the starting position $\boldsymbol{p}_1$. It can be seen that though the starting position searched by DoDTrack has errors, the trajectory can be progressively corrected and restored using EKF.

\begin{figure}[htbp]
    \centering
    \vspace{-3pt}\includegraphics[width=0.75\columnwidth]{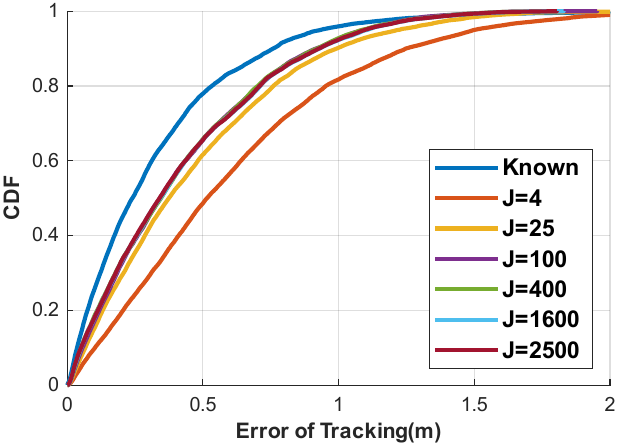}
    \caption{CDF of tracking error.}
    \label{fig: cdf}
    \vspace{-5pt}
\end{figure}

\cref{fig: cdf} shows the cumulative distribution function (CDF) of the tracking error versus the number of grid points $J$. Additionally, the CDF curve corresponding to the case where the starting position $\boldsymbol{p}_1$ is known a prior is provided as a reference and marked in blue. When the number of grid points is very small, the error is relatively large. However, once the number of grid points exceeds a certain threshold, increasing the grid points further yields trivial performance gains. Thus, $J=100$ is a balanced choice between precision and computational resources. The median error for $J=100$ is 0.34 m and the 90\% error is 0.92 m.

For certain trajectories, e.g., \cref{fig: b}, due to measurement errors, the estimated trajectory may deviate from the ground truth even when the starting position is fully known. Perfect knowledge of the starting position may even yield worse results than the proposed method when the starting position is unknown. In summary, the prior knowledge of the starting position results in more accurate tracking result in general, and DoDTrack can still achieve satisfactory tracking accuracy even when the starting position is unknown.

\section{Conclusion}\label{conclusion}
In this paper, we presented DoDTrack, a high-precision indoor tracking system that utilizes the DoDs as the only observations to track active Wi-Fi devices. Unlike conventional DFS-based approaches that require prior knowledge of the transmitter's starting position or complementary information like AoA or ToF, DoDTrack effectively eliminates the carrier frequency offsets by performing conjugate multiplication on signals from receive antennas sharing a common oscillator. By integrating the proposed EKF method and the gradient-descent-based starting position search algorithm, the system achieves rapid and accurate reconstruction of complex trajectories using a single receiver. We believe this work advances indoor device tracking and paves the way for future low-cost tracking and positioning designs through innovative ISAC designs.

\bibliographystyle{IEEEtran}
\bibliography{ref}

\end{document}